# Willingness to Pay and Attitudinal Preferences of Indian Consumers for Electric Vehicles

**Prateek Bansal** (Corresponding author)

Transport Strategy Centre, Department of Civil and Environmental Engineering

Imperial College London, London, UK

prateek.bansal@imperial.ac.uk

**Rajeev Ranjan Kumar**

Production, Operations & Decision Sciences

XLRI Xavier School of Management, Jamshedpur, India

r17005@astra.xlri.ac.in

**Alok Raj**

Production, Operations & Decision Sciences

XLRI Xavier School of Management, Jamshedpur, India

alokraj@xlri.ac.in

**Subodh Dubey**

Department of Transport and Planning

Delft University of Technology, Netherlands

S.K.Dubey@tudelft.nl

**Daniel J. Graham**

Transport Strategy Centre, Department of Civil and Environmental Engineering

Imperial College London, London, UK

d.j.graham@imperial.ac.uk

**Abstract**

Consumer preference elicitation is critical to devise effective policies for the diffusion of electric vehicles (EVs) in India. This study contributes to the EV demand literature in the Indian context by (a) analysing the EV attributes and attitudinal factors of Indian car buyers that determine consumers' preferences for EVs, (b) estimating Indian consumers' willingness to pay (WTP) to buy EVs with improved attributes, and c) quantifying how the reference dependence affects the WTP estimates. We adopt a hybrid choice modelling approach for the above analysis. The results indicate that accounting for reference dependence provides more realistic WTP estimates than the standard utility estimation approach. Our results suggest that Indian consumers are willing to pay an additional US$10-34 in the purchase price to reduce the fast charging time by 1 minute, US$7-40 to add a kilometre to the driving range of EVs at 200 kilometres, and US$104-692 to save US$1 per 100 kilometres in operating cost. These estimates and the effect of attitudes on the likelihood to adopt EVs provide insights about EV design, marketing strategies, and pro-EV policies (e.g., specialised lanes and reserved parking for EVs) to expedite the adoption of EVs in India.

***Keywords*:** Electric vehicle: Willingness to pay; Hybrid choice model; Reference dependence; Indian consumers.

# 1. Introduction

The transportation sector is the third-largest $CO_2$ emitter in India, which accounted for around 11% of the total $CO_2$ emissions in 2016 (Janssens-Maenhout et al., 2017). Continuous increase in sales of fossil fuel vehicles due to growing household income and rapid urbanization creates an alarming situation regarding rising $CO_2$ emissions in India. NITI Aayog, an Indian government policy think tank, predicts that $CO_2$ emissions from the automobile sector may triple by 2030 (NITI Aayog and Rocky Mountain Institute, 2017a).

In the last decade, electric vehicles (EVs) have emerged as the potential alternative to internal combustion engine vehicles (ICEVs) to curb fossil fuel consumption (Abouee-Mehrizi et al., 2020). However, there are many barriers to the consumer adoption of EVs, such as high upfront cost, lower driving range, and lack of trust in EV technology. These challenges become even more pronounced in emerging economies like India due to lower disposable income of consumers, poor charging infrastructure, and lack of awareness about EV technology (Government of India, 2019a; Government of India, 2019b; Vidhi and Shrivastava, 2018). Diffusion of new technologies like EVs is more cumbersome because balancing the interplay between supply and demand becomes a *chicken-and-egg* problem. For instance, consumer demand largely depends on the availability of the charging infrastructure and driving range of EVs, but these supply-side investments are likely to be driven by consumer demand.

To address this *chicken-and-egg* problem, the Indian government plans to address supply-side challenges in EV diffusion through various initiatives, such as the National Electric Mobility Mission Plan (NEMMP) and Faster Adoption and Manufacturing of Electric Vehicles (FAME). NEMMP highlights the Indian government's vision to expedite consumer acceptance and manufacturing of EVs (IEA, 2020). Similarly, the objective of the FAME policy is to boost EV demand by improving the charging infrastructure and providing subsidies. In the first phase of FAME, the government supported 2,78,000 EVs in different forms with a total incentive of 3.43 billion Indian rupees (INR) (~US$47 million). The Indian government launched Phase 2 (FAME-II) in April 2019 with a substantial budget of INR 100 billion (~US$1.36 billion) for the next three years to create a robust eco-system for EVs (Government of India, 2019a). Despite all these supply-side interventions, Indian car buyers are reluctant to buy EVs. In 2019-2020, only 3400 electric cars were sold in India (Venkat, 2020).

This demand pattern suggests that the success of all government initiatives hinges on the preferences of Indian vehicle buyers for EVs. For instance, subsidies on the purchase price might not encourage Indians to buy EVs, but a dense charging infrastructure on highways might drive the EV demand, if Indian consumers are very anxious about the shorter driving range of EVs. Therefore, we investigate the following four research questions to better understand the preferences of Indian car buyers for EVs:

1. Which product (e.g., driving range), service (e.g., charging infrastructure), and policy (e.g., free parking for EVs) attributes determine the preferences of Indian car buyers for EVs?
2. Consumer preferences for EVs are likely to be affected by their latent psychological constructs such as environment-friendliness, hedonic, and symbolic attributes (Schuitema et al., 2013; Skippon and Garwood, 2011). We thus explore – which attitudinal factors affect the EV adoption of Indian consumers?
3. Automakers need to know the willingness to pay (WTP) of Indian car buyers to set the market price and decide the research and development budget to improve EV attributes. To bridge this gap, we estimate – how much are Indian consumers willing to pay to buy EVs with improved attributes?
4. Mabit et al. (2015) and Mabit & Fosgerau (2011) argue that vehicle buyers evaluate utility of alternative fuel vehicles by comparing their attributes against those of ICEV (reference alternative). Considering that the discrete choice experiment design is inherently pivoted around ICEVs and reference-dependent utility results into better explanatory power (Bateman et al., 2009; Kim et al., 2020), we also investigate – do Indian car buyers exhibit *reference dependence* while comparing EVs with ICEVs and how does its inclusion affect WTP estimates?

This study thus contributes to the literature from both empirical and methodological perspectives. From an empirical standpoint, this is the first study in the Indian context that estimates the WTP of Indian consumers for various EV attributes. More specifically, we design a discrete choice experiment and collect stated preferences (SP) of over 1000 Indian consumers for EVs under many scenarios with varying attribute levels. We subsequently analyse the SP data to quantify the effect of product, service, policy, and attitudinal attributes on the preferences of Indian consumers for EVs. Even in other geographical contexts, only a handful of studies have conducted such comprehensive analysis (Ghasri et al. 2019; Jensen et al., 2013; Kim et al., 2014; Nazari et al. 2019).

From a methodological viewpoint, this study analyses the SP data by considering attitudinal characteristics as well as reference dependence for all vehicle attributes using an integrated choice and latent variable (ICLV) model (Ashok et al., 2002; Ben-Akiva et al., 2002). We adopt the curvature-based utility specification to model the reference dependence, which nests the traditional utility specification when the curvature value is one. Such general reference-dependent utility specification and attitudinal characteristics have not been simultaneously considered by any of the three previous studies on reference dependence in eliciting preferences for EVs (Mabit et al., 2015; Mabit & Fosgerau, 2011; Kim et al., 2020).

The remainder of the paper is organised as follows: Section 2 details the contextual literature review. Section 3 summarises experiment design, data collection and sample characteristics. Section 4 presents the ICLV formulation and resulting equations to estimate the WTP. Section 5 discusses the results of ICLV, WTP estimates, and the policy implications of results for practitioners. Section 6 highlights key takeaways and avenues for future research.

## 2. Literature Review

This section has three subsections. We first summarise the related literature on consumer preferences for EVs in different geographical contexts. Subsequently, we discuss the literature on EVs in the Indian context. Finally, we review the modelling approaches adopted by previous studies. We conclude this section with a brief discussion of the identified research gaps.

### 2.1 Studies on EV Preferences

Analysis of consumer preferences for EVs has received considerable attention in the last decade. Table 1 summarises recent studies (2016-20), their study area, sample size, adopted model, and considered attributes. In terms of geographical spread, the majority of studies have been conducted in developed countries such as Australia (Ardeshiri and Rashidi, 2020; Gong et al., 2020), Nordics countries (Noel at al., 2019; Orlov and Kallbekken, 2019), European Countries (Danielis et al., 2020; Hackbarth and Madlener, 2016), Canada (Abotalebi et al., 2019; Ferguson et al., 2018), USA (Cirillo et al., 2017; Sheldon et al., 2017), and South Korea (Choi et al., 2018). Among developing countries, most studies have been conducted in different parts of China (Ma et al., 2019; Qian et al., 2019). We succinctly discuss the attributes considered by these studies.

These studies mainly focus on estimating WTP to improve various product and service attributes of EVs (e.g., driving range and charging infrastructure). It is worth noting that most studies rely on stated preference (SP) data. This is because the choice set considered by a consumer remains unobserved in revealed preference (RP) data, and multi-collinearity between attributes creates inferential challenges (Axsen et al., 2009; Brownstone et al., 2000; Grisolía & Willis, 2016; Hidrue et al., 2011). Moreover, the effectiveness of various policy levers cannot be tested using RP data (Ghasri et al., 2019), which are critical to increasing the market penetration of EVs.

The review suggests that researchers have also explored the impact of several privilege-driven policies on consumer preferences for EVs. These policies range from providing EVs with access to specialised lanes (bus or high occupancy vehicle lane) to relaxing fees associated with public charging, parking, and congestion pricing (Abotalebi et al. 2019; Ferguson et al., 2018; Langbroek et al., 2016; Ma et al., 2019; Wang et al., 2017). Policies like access to specialised lanes for EV drivers has already been implemented in Norway (Figenbaum, 2017). Since China moderates car sales by implementing lottery-based licensing policy and some Chinese cities also manage traffic by restricting vehicles based on license plate numbers, alleviating these purchase and traffic restriction for EV owners can encourage car buyers to purchase EVs (Ma et al., 2019; Qian et al., 2019; Wang et al., 2017).

Many studies also argue that attitudinal factors also affect consumer preferences. These latent attitudes include risk-averse decision-making (Huang and Qian, 2018; Orlov and Kallbekken, 2019), environment-friendliness (Costa et al., 2019; Nazari et al., 2019), trust in technology (Axsen et al., 2016; Ghasri et al., 2019; Kormos et al., 2019), general awareness about EVs (Gong et al., 2020; Lin and Tan, 2017), hedonic and symbolic values (Schuitema et al., 2013; Skippon and Garwood, 2011), social network and norms (Barth et al. 2016; Rasouli and Timmermans, 2016; White and Sintov, 2017).

**2.2 Literature in the Indian Context**

The literature on EV preference elicitation in the Indian context is sparse and limited. Most studies have either used structural equation modelling (SEM), regression/correlation-based analysis or qualitatively analysis.

Digalwar and Giridhar (2015) use SEM to identify the barriers to EV adoption. The authors suggest that better battery technology, increasing awareness, and economic viability are some of the enablers to increase EV adoption. Among the most recent studies, Khurana et al. (2020)

and Navalagund et al. (2020) also adopt SEM to understand the roles of attitude in EV adoption. Khurana et al. (2020) use a sample of 214 respondents from Delhi, Mumbai and Pune, and Navalagund et al. (2020) collect survey response of 384 respondents from Pune to evaluate the effect of self-image, perceived economic advantages, knowledge of EVs, and environmental concerns on the consumer preferences for EVs.

Bhalla et al. (2018) apply correlation analysis on the survey data of 233 respondents to study the factors influencing the consumers' perception towards EVs. They find that environmental concern and trust in technology are the key factors that influence consumers' choices for EVs. Motwani and Patil (2019) use a regression model to analyse the preferences of Indian consumers. Based on survey responses from 345 vehicle owners, they find that mobility and recharging characteristics are the most significant determinants of consumers' preferences for EVs.

Among qualitative studies, Bansal and Kockelman (2017) conduct a survey of experts and find that poor charging infrastructure, lack of public awareness about the benefits of EVs, and higher upfront cost are the major obstacles in the adoption of EVs in India. Vidhi and Shrivastava (2018) analyse shared electric mobility services and recommend various privilege-driven policies to encourage car owners to leave their cars and adopt shared electric services.

A few studies have focused on supply-side interventions to improve EV adoption in India. Awasthi et al. (2017) use genetic algorithm and particle swarm optimization to design the charging infrastructure in Allahabad, India. Kumar et al. (2015) suggest that the vehicle-to-home scheme could help to increase the penetration of EVs in India by reducing the payback period of EV owners. Their main idea is to utilise the parked EVs to store electrical energy. This strategy would help in levelling the peak of the Indian power grid, and the cost benefits would be transferred to EV owners.

**2.3 Modelling Approaches for WTP Estimation**

In terms of modelling approaches to estimate the WTP, the majority of studies rely on random-utility-theory-based multinomial logit model (MNL) and its extensions (Costa et al., 2019; Orlov and Kallbekken, 2019). For instance, researchers have employed nested logit (Huang and Qian, 2018; Nazari et al., 2019) and multinomial probit model (Higgins et al., 2017) to capture a rich substitution pattern between alternatives. Mixed-MNL (Choi et al., 2018; Cirillo et al., 2017) and latent class MNL (LC-MNL) (Ardeshiri and Rashidi, 2020; Axsen et al., 2016; Ferguson et al., 2018; Kormos et al., 2019) are generally adopted to capture observed and unobserved heterogeneity in preferences of car buyers. LC-MNL is preferred over Mixed-MNL for a policy analysis because it provides more interpretable behavioural insights about the sources of heterogeneity. LC-MNL identifies consumer segments based on their attitudes and socio-demographic characteristics, while jointly estimating segment proportions and segment-specific WTP (Ardeshiri and Rashidi, 2020; Hidrue et al., 2011). For instance, Ferguson et al. (2018) use LC-MNL to identify four classes based on Canadian household's mindset.

LC-MNL can thus incorporate attitudinal characteristics for market segmentation, but it has three major limitations that ICLV can address. First, to be able to use the parameter estimates of LC-MNL for forecasting, the analyst needs to know the attitudinal response of consumers. However, by mapping attitudinal indicators, latent variables, and socio-demographics in the SEM component, forecasting with ICLV does not require attitudinal indicators (Kamargianni et al., 2015). Of course, this argument pivots on the temporal stability of the attitudes. In the absence of any extreme event between the data collection and prediction stage, we can safely assume attitudes to be temporally stable (Sheeran and Abraham, 2003). Second, including too many indicators to represent the latent consumer behaviour in the segmentation model can make the LC-MNL empirically fragile (i.e., poor interpretability and numerical issues in the estimation) due to multi-collinearity and explosion of the parameter space (Bhat and Dubey, 2014). For instance, including response to nine five-Likert scale questions leads to thirty-six heavily correlated explanatory variables. ICLV can handle this challenge by mapping attitudinal characteristic to a lower-dimensional space of latent variables (Ben-Akiva et al., 2002). Third, since attitudinal responses are just a proxy for the underlying attitudes and are observed with measurement error, using them directly as explanatory variables can lead to biased estimates (Ashok et al., 2002). ICLV framework provides a seamless way to simultaneously derive the underlying attitudes from the measurements and correct for such measurement errors.

In addition, ICLV offers a mathematical framework to underpin the sources of preference heterogeneity (Ashok et al., 2002; Ben-Akiva et al., 2002; Bhat and Dubey, 2014). By interacting latent variables with observed attributes in the indirect utility, one can derive various insights about the heterogeneity in willingness to pay of consumers with different attitudinal characteristics. ICLV could unveil the black box of heterogeneity because latent variables can be well-labelled with attitudinal characteristics. Using an analytical approach and Monte Carlo simulation, Vij and Walker (2016) illustrate that ICLV can also improve predictive abilities of the reduced-form choice models, correct for omitted variable bias, and reduce the variation of parameter estimates in many situations. Readers are referred to Vij and Walker (2016) to learn more about the importance of ICLV for policymakers.

There are proven empirical benefits of ICLV in understanding various types of choice behaviour – food choice (Alemu and Olsen, 2019), route choice (Alizadeh et al. 2019), freight transport behaviour (Bergantino et al., 2013), Pedestrian crossing behaviour (Cantillo et al. 2015), preferences for autonomous vehicles (Sharma and Mishra, 2020), residential location choice (Kitrinou et al., 2010), shared mobility choices (Li and Kamargianni 2020), preferences for online shopping activity (Schmid and Axhausen, 2019). However, applications of ICLV in EV preference modelling are limited (Ghasri et al. 2019; Jensen et al., 2013; Kim et al., 2014; Nazari et al. 2019).

Ghastri et al. (2019) adopt ICLV to measure the perceived advantages of EVs in New South Wales, Australia. They include three attitudinal variables in SEM to capture consumer perception regarding the effect of EVs on the environment, EV design, and safety benefits of EVs. Jensen et al. (2013) estimate an ICLV model to incorporate consumers' concerns about environment, interest in technology, and notion of car as a status symbol while investigating the effect of consumers' experience of riding an EV on their preferences. Kim et al. (2014) develop an ICLV-based purchase intention model, which simultaneously estimates the effect of vehicle attributes, latent attitudes (environment, battery, and innovation), and social influence on EV preferences. Nazari et al. (2019) sequentially estimate SEM and nested logit model. Whereas SEM determines latent constructs associated with green travel patterns, nested logit model estimates fuel type preferences.

**TABLE 1** Recent studies (2016-2020) relying on discrete choice analysis to understand EV preferences.

| Authors (in alphabetical order) | Models | Country | Sample size | Product and service attributes | | | | | | | | | Policy attributes | | | | Attitudinal attributes | | | | |
|---|---|---|---|---|---|---|---|---|---|---|---|---|---|---|---|---|---|---|---|---|---|
| | | | | E-Charging Infra | Fast charging time | Slow charging time | Driving range | Upfront cost | Operating cost | Emissions | Vehicle size / type | Acceleration / Speed | Special lane access | Free parking | Free charging | Free toll roads | Risk averse | Pro-environment | Pro-technology | EV knowledge | Social effect |
| Abotalebi et al. (2019) | LC-MNL | Canada | ~11,500 | √ | √ | √ | √ | √ | √ | √ | √ | √ | √ | √ | | √ | | √ | | √ | |
| Ardeshiri and Rashidi (2020) | LC-MNL | Australia | 1,180 | √ | | | | | | | | | | | | | | | | √ | |
| Axsen et al. (2016) | LC-MNL | Canada | 1,848 | | √ | √ | √ | √ | √ | | | | | | | | | √ | √ | | |
| Choi et al. (2018) | Mixed-MNL | South Korea | 1,002 | √ | | | | √ | √ | | | | | | | | | | | | |
| Cirillo et al. (2017) | Mixed-MNL | USA | 456 | | | | √ | √ | √ | | √ | | | | | | | | | | |
| Costa et al. (2019) | MNL | Italy | 278 | | | | | √ | | √ | √ | | | | | | | √ | | | |
| Danielis et al. (2020) | Mixed-MNL | Italy | 996 | √ | √ | | √ | √ | √ | | | | | √ | | | | √ | | √ | |
| Ferguson et al. (2018) | LC-MNL | Canada | 17,953 | √ | √ | √ | √ | √ | √ | | √ | √ | √ | √ | | √ | | | | | |
| Ghasri et al. (2019) | ICLV | Australia | 1,076 | √ | √ | √ | √ | √ | √ | | √ | | √ | √ | | | | √ | √ | | √ |
| Gong et al. (2020) | LC-MNL | Australia | 1,076 | √ | √ | √ | √ | √ | √ | | √ | | √ | √ | | | | | | √ | √ |
| Hackbarth and Madlener (2016) | LC-MNL | Germany | 711 | √ | | | √ | √ | √ | √ | √ | | √ | √ | | | | √ | √ | | |
| Higgins et al. (2017) | MNP | Canada | 20,520 | √ | √ | √ | √ | √ | √ | | √ | √ | √ | √ | | √ | | | | | |
| Huang and Qian (2018) | NL | China | 348 | √ | | | | √ | √ | √ | | | √ | | | √ | √ | | | | √ |
| Kormos et al. (2019) | LC-MNL | Canada | 2,123 | √ | √ | √ | √ | √ | | | | | | | | | | √ | √ | | |
| Langbroek et al. (2016) | Mixed-MNL | Sweden | 294 | | | | √ | √ | | | | | √ | √ | √ | | | | | | |
| Lin and Tan (2017) | Ordered Probit | China | 958 | √ | | | | √ | | | | | | | | | | √ | | √ | |
| Ma et al. (2019) | Mixed-MNL | China | 1,719 | √ | √ | | √ | √ | | | | | √ | √ | √ | √ | | | | | |
| Nazari et al. (2019) | SEM and NL | USA | 39,250 | √ | | | | | | | | | √ | | | | | √ | | | |
| Noel at al. (2019) | Mixed-MNL | Nordics | 4,105 | | √ | √ | √ | √ | | | | √ | | | | | | | √ | | |
| Orlov and Kallbekken (2019) | MNL | Norway | 1,093 | | | | | √ | √ | | √ | | | | | | √ | √ | | √ | √ |
| Qian et al. (2019) | Mixed-MNL | China | 1,076 | √ | √ | √ | √ | √ | | | | | | | | | | | | | |
| Rasouli and Timmermans (2016) | Mixed-MNL | Netherlands | 726 | √ | √ | √ | √ | √ | √ | | | √ | | | | | | | | | √ |
| Sheldon et al. (2017) | Mixed-MNL | USA | 1,261 | √ | | | √ | √ | √ | | √ | | √ | | | | | √ | | | |
| Wang et al. (2017) | Mixed-MNL | China | 247 | | | | √ | √ | | | | | √ | √ | √ | √ | | | | | |
| **Present Study** | **ICLV** | **India** | **1,021** | √ | √ | √ | √ | √ | √ | | | | √ | √ | | | √ | √ | √ | √ | √ |

**Note:** MNL: Multinomial Logit Model; MNP: Multinomial Probit; NL: Nested Logit; LC: Latent Class; ICLV: Integrated Choice and Latent Variable; SEM: Structural Equation Model.

Based on the above discussion, we have identified two main research gaps. First, despite the importance of EVs in the Indian context, very few studies have investigated the factors affecting the demand for EVs. The literature review indicates that most of the previous studies in the Indian context are either speculative or rely on a small sample, and none of them estimates the WTP of Indian car buyers for EV attributes. Second, whereas only a few EV preference studies account for latent attitudinal attributes, all of them ignore *reference dependence* behaviour. To bridge these gaps in the literature, we present the first consumer behaviour analysis that estimates the WTP of Indian car buyers for various EV attributes. We conduct this analysis using an ICLV model that simultaneously accounts for various behavioural aspects, including latent attitudes and *reference dependence*.

## 3. Experiment Design and Data Collection

In this section, we discuss the selection of attribute levels for experiments, followed by experiment design, data collection procedure, and sample summary statistics.

### 3.1 Selection of Attribute Levels

In the choice experiment, we ask respondents to choose between ICEV and EV in hypothetical scenarios. It is worth noting that we only provide battery EV (BEV) as an alternative in the experiment due to the non-existence of plug-in EVs in India (see Chan (2007) for the distinction between different EV types). In designing choice scenarios, we consider three types of attributes –product, service, and policy. The choice of attributes is based on the literature review and relevance in the Indian context. We obtain the information related to product attributes to create attribute levels in experiment design from a web portal – CarDekho, India's leading car search engine (CarDekho, 2020).

Table 2 summarises levels of all considered attributes in the choice experiment. The included product attributes are on-road purchase price, running (operating) cost, and driving range after full charging. We first ask respondents about the price of the ICEV that they recently bought or would consider to buy. This reported price is shown as the on-road price of ICEV in the experiment. We consider three levels for EV on-road price – 30%, 45%, and 60% higher than the ICEVs price. The resulting EV price ranges are in line with the upfront cost of available BEVs in the Indian market, such as Mahindra-electric and Tata Tigor. It is worth noting that when attributes are pivoted on the reference alternative, such designs are known as *pivot*

*designs* in the literature (Bansal and Daziano, 2018; Rose et al., 2008) and are commonly used in EV preference elicitation (Hidrue et al., 2011).

Considering that gasoline price in India is around INR 75 per litre (GlobalPetrolPrices, 2020) and fuel economy of the available hatchback and sedan variants is between 15 to 25 km per litre, we consider three levels of ICEV operational cost – INR 3, INR 4, and INR 5 per kilometre. Based on the running performance of existing EV models, the operating cost of EV is also assumed to have three levels – INR 0.5, INR 1.0, and INR 1.5 per kilometre. We also ask respondents about their weekly travel distances. Using fuel cost and distance information, we compute and use weekly operating cost as an attribute in the utility specification.

**TABLE 2** Attributes and levels in the discrete choice experiment.

| Attributes | Units | ICEV 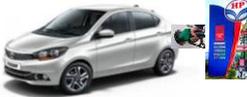 | EV 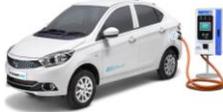 |
|---|---|---|---|
| **Product attributes** | | | |
| On-road purchase price | INR | Specified by the respondent | 1. 30% higher than the cost of ICEV<br>2. 45% higher than the cost of ICEV<br>3. 60% higher than the cost of ICEV |
| Running cost | INR/km | 1. INR 3.0/km<br>2. INR 4.0/km<br>3. INR 5.0/km | 1. INR 0.5/km<br>2. INR 1.0/km<br>3. INR 1.5/km |
| Driving range | kms | 1. 600 kms<br>2. 800 kms | 1. 150 kms<br>2. 200 kms<br>3. 250 kms |
| **Service attributes** | | | |
| Slow charging time | Hours | Not applicable | 1. 6 hours<br>2. 8 hours<br>3. 10 hours |
| Fast charging time | Minutes | 1. 5 minutes<br>2. 10 minutes | 1. 30 minutes<br>2. 60 minutes<br>3. 90 minutes |
| Availability of fast charging stations | Distance between stations | 1 km (within city) | 1. 3 km (within city)<br>2. 5 km (within city)<br>3. 7 km (within city) |
| **Policy attribute** | | | |
| Reserved parking | | No | 1. Yes<br>2. No |
| Specialised lanes in congested areas | | No | 1. Yes<br>2. No |

**Note:** INR: Indian Rupees; km: kilometer.

We also consider driving range as an additional product attribute. Many earlier studies have identified it as the main barrier to EV adoption (Lim et al. 2015; Skippon et al. 2016; Melliger et al., 2018). In the experiment, we set three levels of driving range of EVs – 150 kilometres,

200 kilometres, and 250 kilometres. These values are aligned with the driving range of the existing EV models in hatchback and sedan segments in India. Considering that these small vehicle size segments constitute more than 70% (see Indian Automobile Market report) of cars sold in India (mostly due to budget constraints), we restrict the upper limit of driving range to 250 kilometres. Moreover, we also did not provide larger values of the driving range to avoid the hypothetical bias, which is inherently a concern in stated preference experiments. We compute the driving range of ICEV based on the fuel tank capacity (~40 litres) and mileage of the vehicle (~15 to 20 kms/liter) and vary it at two levels – 600 kilometres and 800 kilometres.

Among service-related attributes, we consider fast charging time, slow charging time and density of fast charging stations within the city. These attributes fully represent the level of service (Quian et al. 2019; Gong et al. 2020). Since charging points at workplace or home typically use slow charging technology, 6 to 10 hours are required to fully charge an EV (Liu, 2012). Therefore, we consider three levels for EV slow charging – 6 hours, 8 hours, and 10 hours. Slow charging is not applicable for ICEVs. Whereas fast charging time of EVs has three levels – 30 minutes, 60 minutes, and 90 minutes, analogous refuelling time for ICEVs varies at two levels – 5 minutes and 10 minutes. Given that the Indian government aims to install at least one publicly accessible charger within a square grid of 3 kilometres in selected cities under FAME-II scheme (Government of India, 2019b), we assume three levels of the density of fast chargers – at every 3 kilometres, 5 kilometres, and 7 kilometres. As per Indian government notification, the minimum distance between two fuel stations should be between 300 and 1000 meters, depending on the locality (Ministry of Road Transport and Highways, 2020). Therefore, fuel stations are considered to be available at every kilometre within the city.

Based on the Government's ongoing considerations for privilege-driven policies to promote EV adoption, we consider two policies – reserved parking (Dogra, 2019) and access to specialised lanes for EVs (DHNS, 2020). Previous studies have also explored the effect of such policies on consumer preferences (Ma et al., 2019; Wang et al., 2017).

**3.2 Design of Choice Experiment**

Discrete choice experiments are generally created using an efficient or a randomised design. Efficient designs have become popular in the last decade (Kessels et al., 2006; Rose and Bliemer, 2009). However, Walker et al. (2018) illustrate that the analyst needs to be cautious about the choice of priors on model parameters in these designs. The efficient designs can even perform worse than traditional randomised designs in case of ill-chosen priors (Walker et al., 2018). Since there are no existing studies in the Indian context, we do not have any good basis to select priors on model parameters. Therefore, we adopt randomised experiment design, while ensuring that *comparison relations* between attributes of ICEV and EV hold (e.g., purchase price of ICEVs is less than that of EVs) and price of EV is pivoted on the reference price reported by the survey respondent. We design 24 scenarios using these attribute levels and randomly select 3 of them for each respondent. Since attributes have 1, 2, or 3 levels, 24 choice situations also ensure attribute level balance in design.

We inform survey participants that EVs have zero tailpipe emission. They have a higher purchase price but lower operating cost, lower driving range than their counterpart ICEVs. Moreover, we also conveyed information about the differences between slow and fast charging of EVs, and the required infrastructure for both. To reduce the cognitive effort of the respondents, we provide images of both EV and ICEV with fuel pump/charging unit. Figure 1 shows a sample of the choice situation presented to the survey participants.

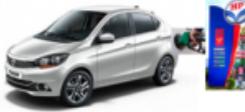

**FIGURE 1** A sample of the choice situation

### 3.3 Data Collection and Summary Statistics

We collect data on EV preferences of Indian vehicle buyers between March and June 2020 using a nationwide stated preference survey. In addition to the choice experiment, we ask respondents about their socio-demographic characteristics and ordinal indicator questions to capture their latent attitudes. We create a web-based survey on the Qualtrics platform and recruit respondents using a snowball sampling procedure. Specifically, we target a training programme related to supply chain and transportation, that was conducted in March 2020 in

XLRI (one of the top Business schools in India) and received PAN India participation. We first identify potential respondents from the training program and subsequently use their referrals to identify other respondents. Participation was voluntary, and participants were informed about the anonymity and confidentiality of the survey data. Similar data collection process has been followed in previous studies (Jiang et al., 2019; Pepermans, 2014).

We first acquire 110 responses in a pilot study and estimate a multinomial logit model. This preliminary analysis provides the initial evaluation of the experiment design. After slightly modifying the design based on the results of the pilot study, we collect the final sample from a wide range of urban areas in India. Out of 2,176 visitors, 47.38% (i.e., 1,031 respondents) complete the survey. Figure 2 shows the regional distribution of these 1031 respondents. The sample covers multiple states with the maximum number of respondents from Maharashtra (186), followed by Delhi (155), Karnataka (145), Tamil Nadu (127), Telangana (117), Gujarat (88) and West Bengal (74). In terms of spread across cities, a maximum number of responses are from Delhi, Mumbai, Bangalore, Hyderabad, Chennai, Ahmedabad, and Kolkata (in decreasing order), which are also the top seven most populated cities in India. After removing responses with short completion time (below 40% of the median response time), the sample has 1,021 respondents for the final analysis (see Table 3 for sample summary).

**TABLE 3** Summary of the sample demographics (N=1,021)

| Characteristics | Percentage | Characteristics | Percentage |
|---|---|---|---|
| *Marital status* | | *Education level* | |
| Married | 55.1 | Below bachelor's degree | 11.7 |
| Single | 44.0 | Bachelor's degree | 41.6 |
| Others | 0.9 | Master's or higher degree | 46.7 |
| *Age* | | *Annual household income (INR)* | |
| 18-30 | 48.5 | Less than 5,00,000 | 20.8 |
| 31-40 | 37.1 | 5,00,001-10,00,000 | 25.1 |
| 41-50 | 10.2 | 10,00,001-20,00,000 | 35.8 |
| 51-60 | 3.1 | 20,00,001-30,00,000 | 12.5 |
| More than 60 | 1.1 | More than 30,00,000 | 5.8 |
| *Family size* | | *Car ownership* | |
| 2 or fewer | 8.7 | No car | 31.7 |
| 3 | 24.4 | 1 car | 47.3 |
| 4 | 33.3 | 2 cars | 16.5 |
| 5 or more | 33.6 | 3 cars or more | 4.5 |
| *Number of children* | | *Employment type* | |
| 0 | 68.6 | Private sector | 73.4 |
| 1 | 20.8 | Unemployed | 10.2 |
| 2 | 4.5 | Self-employed | 8.6 |
| 3 or more | 6.1 | Government sector | 7.8 |
| *Gender* | | | |
| Female | 23.6 | | |
| Male | 76.4 | | |

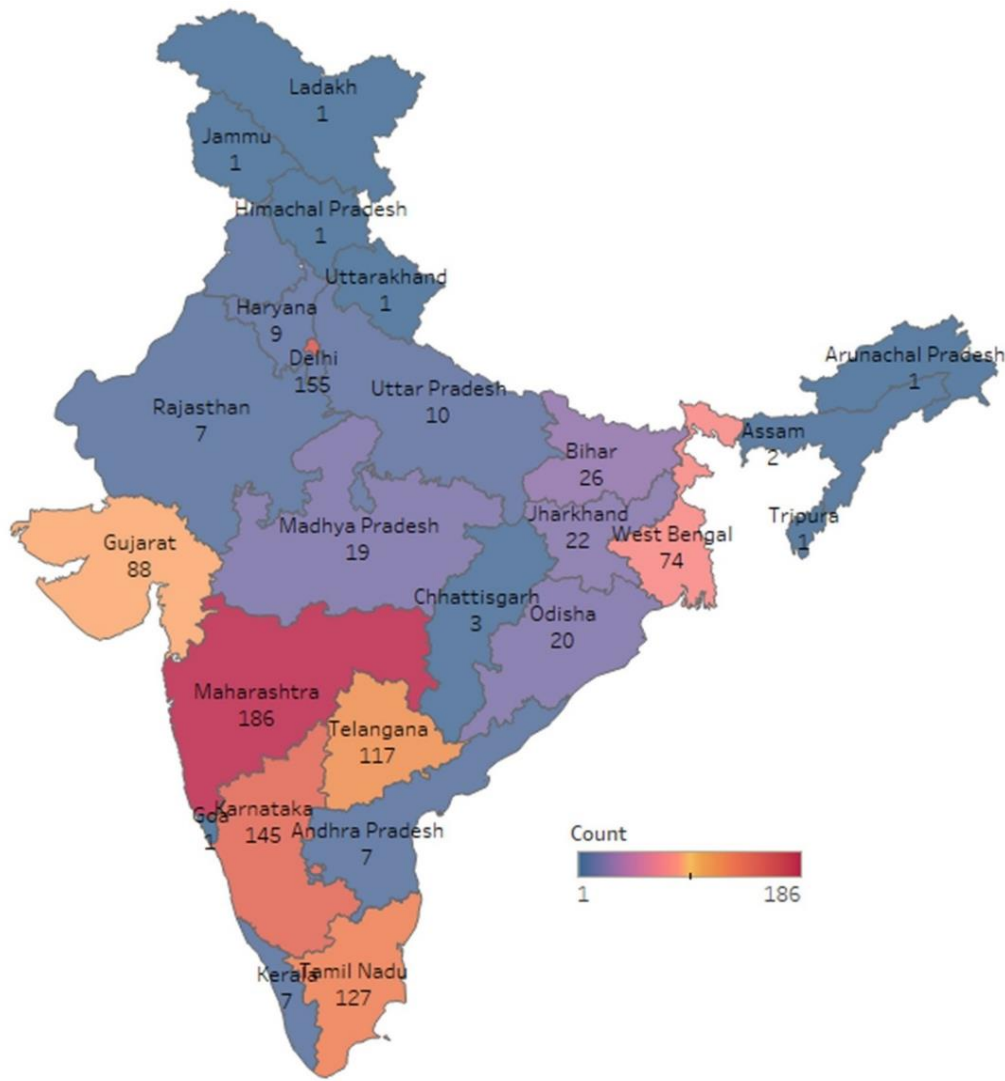

**FIGURE 2** Spatial distribution of the respondents (N=1,031).

It is worth noting that the target population of this study constitutes car owners or potential car buyers in India. Since demographics for such specific population segment are difficult to obtain for developing countries like India due to the scarcity of publicly available disaggregate census data (Bansal et al., 2018), representativeness of the sample is challenging to evaluate. The higher sample proportion of some demographic segments (e.g., high income, high education, and private sector employees) is a consequence of better access to internet facilities and more exposure to the English language to these groups. However, this is not a concern because individuals from these demographic groups are likely to be car owners or potential car buyers who might be interested in EVs.

We measure the attitudinal indicators at a five-point Likert scale (one being strongly disagreed and five being strongly agreed) and summarise the probability mass function of each indicator in Table 4. Through these indicators, we capture the latent attitudes of a respondent to be a climate-doubter (Ind01 to Ind03, i.e. who does not believe in the impact of EVs and human behaviour on climate), an EV-tech believer (Ind04 to Ind06, i.e. who trusts EV technology), and an early adopter (Ind07 to Ind09, i.e. who is likely to adopt EVs at low market penetration). The mean score of above 4 for all climate-related indicators (i.e., Ind01 to Ind03) shows that Indians are concerned about the environment, and they think that EVs can help in battling climate change. The scores of indicators related to EV technology (i.e., Ind04 to Ind06) illustrate that around 45% of respondents do not trust EV technology and mistrust discourages them from purchasing EVs. Around 70-75% of respondents agree that limited service support and inadequate charging infrastructure discourage them from purchasing EVs (see Ind07 and Ind08). We also capture the knowledge of respondents about EVs and social network effects using Ind10 and Ind11. The scores of Ind10 indicate that lack of EV knowledge among Indians is a concern because 61% of respondents have very little awareness or only heard about EVs. Moreover, social network effect is pronounced among Indian consumers as around 43% of respondents are likely to adopt EVs if their friends purchase an EV (see Ind11).

**TABLE 4** Descriptive statistics of the Indicators

| Indicator statements on Likert scale (strongly disagree = 1, strongly agree = 5, if not specified) | 1 | 2 | 3 | 4 | 5 | Mean |
|---|---|---|---|---|---|---|
| *Ind01*: We can reduce climate hazard by changing our behaviour. | 4% | 2% | 10% | 42% | 42% | 4.16 |
| *Ind02*: I am concerned about the influence of human behaviour on climate change. | 2% | 2% | 11% | 53% | 32% | 4.12 |
| *Ind03*: EVs can help in battling climate change. | 2% | 4% | 13% | 44% | 37% | 4.10 |
| *Ind04*: I do not want to take a risk by purchasing an EV since I know little about it. | 9% | 35% | 25% | 25% | 6% | 2.83 |
| *Ind05*: I do not trust EV technology. | 6% | 24% | 25% | 35% | 10% | 3.18 |
| *Ind06*: EV is not a proven technology, which discourages me from purchasing. | 6% | 25% | 24% | 35% | 11% | 3.20 |
| *Ind07*: Limited service support for EVs in India discourages me from purchasing. | 2% | 7% | 17% | 47% | 27% | 3.89 |
| *Ind08*: Not a good idea to purchase EV due to limited charging infrastructure. | 2% | 11% | 16% | 48% | 23% | 3.78 |
| *Ind09*: It is not a good idea to purchase an EV due to low resale value. | 5% | 23% | 32% | 31% | 9% | 3.15 |
| *Ind10*: Knowledge about EVs (1= never heard to 5 = know all). | 2% | 20% | 41% | 33% | 4% | 3.16 |
| *Ind11*: If my friend buys EV, my chance of buying it will increase. | 7% | 17% | 33% | 36% | 7% | 3.19 |

## 4. Modelling Approach

We use an Integrated Choice and Latent Variable (ICLV) model (Ben-Akiva et al. 2002). This model has two components – a discrete choice model (DCM) and SEM. Figure 3 shows the ICLV modelling framework in the context of this study. The mathematical details of each component and WTP computations are discussed below.

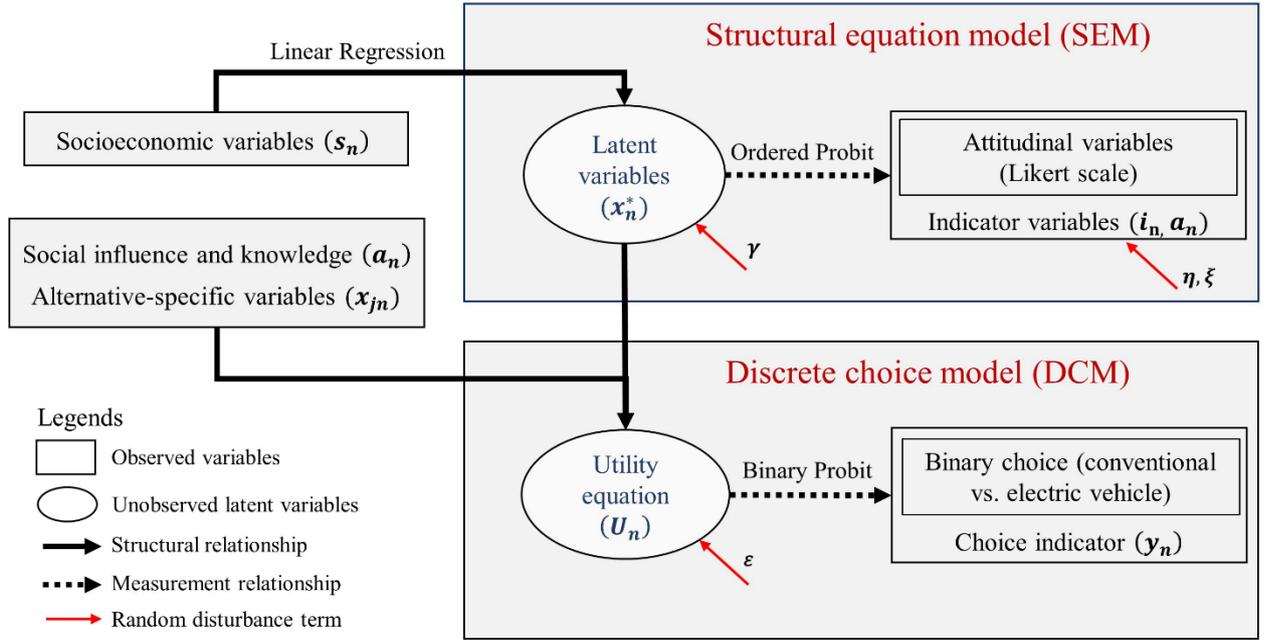

**FIGURE 3** Framework of Integrated Choice and Latent Variable (ICLV) model in the context of this study (adapted from Sharma & Mishra, 2020).

### 4.1 Discrete Choice Model (DCM)

Based on various vehicle-specific attributes, we ask respondents to choose one of two alternatives ($j \in \{EV, CV\}$, $J = 2$) in choice experiments. We consider that vehicle buyer $n$ is a rational decision-maker and she chooses EV if the random utility derived from EV is higher than ICEV, and vice versa. Equations 1 and 2 are the vector and scalar representation of the structural equation, and Equation 3 is the measurement equation for the DCM component. In Equation 1, the indirect utility vector $U_n$ is the function of vehicle-specific attributes $x_n$ (e.g., purchase price and operating cost), latent variables $x_n^*$, and indicator variables $a_n$, and a zero-mean normally-distributed idiosyncratic error term $\varepsilon_n$ with a variance-covariance matrix $\Omega_\varepsilon$. $\varepsilon_n$ captures the factors considered by the vehicle buyer, but unobserved to the researcher. In Equation 3, we present the measurement model, where $y_{jn}$ is the choice indicator. The resulting choice model is a binary Probit model.

$$U_n = f(x_n, x_n^*, a_n, \varepsilon_n;\ \beta, \alpha, \delta, \varphi, \Omega_\varepsilon) \tag{1}$$

$$U_{jn} = \sum_{l=1}^{L} \beta_l \left( x_{jnl} - x_{jnl}^{ref} \right)^{\alpha_l} + \sum_{r=1}^{R} \delta_{jr}\, x_{nr}^* + \sum_{r=1}^{R}\sum_{l=1}^{L} \varphi_{rl}\, x_{nr}^*\, x_{jnl} + \sum_{i=1}^{I}\sum_{o=1}^{O-1} \vartheta_{jio}\, a_{nio} + \varepsilon_{jn} \tag{2}$$

$$y_{jn} = \begin{cases} 1 & \text{if } U_{jn} > U_{in}\ (\forall i \neq j) \\ 0 & \text{Otherwise} \end{cases} \tag{3}$$

We detail the functional form of the indirect utility in Equation 2, where $U_{jn}$ is the utility that decision-maker $n$ derives from choosing alternative $j$. Following the prospect theory (Tversky and Kahneman, 1991), we consider that vehicle buyers evaluate vehicle attributes $x_{jnl}$ with respect to a reference level $x_{jnl}^{ref}$ (i.e., reference dependence), and their marginal utility of gains and losses decreases with the increase in the deviation from the reference level (i.e., diminishing sensitivity). We incorporate the reference dependence by considering ICEV attribute value as the reference level for the EV attribute and diminishing sensitivity is controlled by the estimable *curvature* parameter $\alpha_l$ for attribute $l$. Prospect theory also argues that the utility function is steeper for losses than for gains (Hardie et al., 1993), but we could not incorporate such loss aversion in our analysis because the choice experiment is designed in a way that either EV is superior to ICEV for an attribute or vice versa. For instance, driving range of EV is always lower and purchase price of EV is always higher than those of ICEV at all choice occasions.

Apart from vehicle-specific attributes $\boldsymbol{x_n}$, we think that three attitudes ($x_{nr}^*$, $R = 3$) – consumer's notion about the significance of EVs in the battle against climate change (i.e., climate-doubters), belief in EV technology (i.e., EV-tech believers), and tendency to adopt technologies at low market penetration (early-adopters) – can potentially determine her latent intentions to purchase an EV over ICEV. Since the effect of vehicle-specific attributes on the preference of a vehicle buyer can vary based on her attitudinal characteristics, we also incorporate the interaction of vehicle-specific attributes with latent variables in the utility equation. These interactions also enable an analyst to capture the heterogeneity in the marginal utility of alternative-specific attributes. We control for ordinal indicators for EV knowledge and social network effect after converting them in dummy variables ($a_{nio}$, $I = 2, O = 5$).

For identification, we use ICEV alternative as the base for each nominal variable when introducing alternative-specific constants and covariates that do not vary across alternatives. Since only the covariance matrix of the error differences is estimable and the covariance matrix needs to be scaled for normalization, we impose traditional restrictions on $\boldsymbol{\Omega_\varepsilon}$ for the point identification (Bhat and Dubey, 2014). Since the covariance matrix is of size $2 \times 2$, we end up with no estimable parameter for $\boldsymbol{\Omega_\varepsilon}$ after applying identification conditions.

## 4.2 Structural Equation Model (SEM)

In the SEM component, we first specify the structural relationship between latent variables $x_n^*$ and demographic characteristics $s_n$ using a trivariate linear regression model where idiosyncratic error term $\gamma_n$ follows a trivariate normal distribution with mean $\mathbf{0}$ and variance-covariance matrix $\Omega_\gamma$ (see Equation 4). To ensure identification, we only estimate error correlation matrix $L_\gamma$ (Bhat and Dubey 2014; Stapleton, 1978).

$$x_n^* = \Pi s_n + \gamma_n \tag{4}$$

$$i_{nrk}^* = \tau_{0rk} + \tau_{rk} x_{nr}^* + \eta_{nrk} \tag{5}$$

$$a_{ni}^* = \zeta_{0i} + \sum_{r=1}^{R} \zeta_{ri} x_{nr}^* + \xi_{ni} \tag{6}$$

$$i_{nrk} = \begin{cases} 1 & i_{nrk}^* \leq \psi_{rk1} \\ 2 & \psi_{rk1} < i_{nrk}^* \leq \psi_{rk2} \\ \vdots & \\ O & i_{nrk}^* > \psi_{rk(O-1)} \end{cases} \tag{7}$$

$$a_{ni} = \begin{cases} 1 & a_{ni}^* \leq \mu_{i1} \\ 2 & \mu_{i1} < a_{ni}^* \leq \mu_{i2} \\ \vdots & \\ O & a_{ni}^* > \mu_{i(O-1)} \end{cases} \tag{8}$$

To measure latent variables, we ask three indicator questions for each of three latent variables with responses at a five-point Likert-scale. We use a multivariate ordered Probit model to map indicators with latent variables. Structural and measurement parts of ordered Probit models are presented in Equations 5 and 7 ($r = 1\ to\ 3; k = 1\ to\ 3$), where idiosyncratic error term $\eta_{nrk}$ in Equation 5 follows an independent standard normal distribution for identification (McKelvey and Zavoina, 1975). Moreover, $\tau_{0rk}$ is an intercept, $\psi_{rk}$ is a 4× 1 vector of thresholds, and $i_{nrk}$ is the indicator for the ordinal outcome category chosen by the person $n$ for $k^{th}$ ordinal indicator variable corresponding of $r^{th}$ latent variable. For identification of constants $\tau_{0rk}$, $\psi_{rk1}$ is set to zero for all $r$ and $k$ (Bhat and Dubey, 2014). To ensure that model results can be reused on a different dataset which do not have information on indicators, we also map knowledge and social network effect indicators $a_n$ to all three latent variables using similar ordered Probit specification, as shown in Equations 6 and 8 ($i = 1\ to\ 2$). Following the identification restrictions for the ordered Probit model, $\mu_{i1}$ is set to zero and $\xi_{ni}$ is assumed to follow an independent standard normal distribution for all $i$. Readers can refer to Bhat and Dubey (2014) for theoretical justification behind the considered identification conditions.

### 4.3 Estimation and Willingness to Pay

To estimate the ICLV model, we adopt a composite marginal likelihood (CML) approach. Instead of directly maximising the joint likelihood of the choice and ordinal outcomes, the CML approach maximises an easier-to-compute surrogate objective function which constitutes lower-dimensional marginal likelihoods. More specifically, the surrogate function compounds probabilities of the choice outcome with an ordinal indicator and probabilities of each pair of ordinal indicators. Under typical regularity conditions, the CML estimator is consistent and asymptotically Gaussian (Xu and Reid, 2011). We omit mathematical details of the surrogate objective function for brevity. Interested readers can refer to Section 3.1 of Bhat and Dubey (2014) for the details of estimation and inference of the ICLV model parameters using the CML approach. We write our code in Gauss, a matrix programming language, to estimate ICLV using the CML approach.

The willingness to pay (WTP) of a decision-maker for an attribute of an alternative is the ratio of the marginal utility of the attribute to the marginal utility of its purchase price (see Equation 9). In the case of linear-in-parameter utility specification, WTP turns out to be the ratio of the attribute coefficient to the purchase price coefficient (Daly et al., 2012). However, our utility specification in Equation 2 is non-linear-in-parameters. We present the partial derivative of utility $U_{jn}$ relative to an attribute $x_{jnl}$ in Equation 10. Similar to previous studies, we use logarithmic of range to ensure that marginal utility of range decreases with its magnitude (see Dimitropoulos et al., 2013 for a meta-analysis). Thus, the partial derivative of the utility relative to the driving range (denoted by RNG) is different, which we present in Equation 11.

$$WTP = \frac{\frac{\partial U_{jn}}{\partial x_{jnl}}}{\frac{\partial U_{jn}}{\partial Price_{jnl}}} \tag{9}$$

$$\frac{\partial U_{jn}}{\partial x_{jnl}} = \beta_l \alpha_l \big( x_{jnl} - x_{jnl}^{ref} \big)^{\alpha_l - 1} + \sum_{r=1}^{R} \varphi_{rl}\, x_{nr}^* \tag{10}$$

$$\frac{\partial U_{jn}}{\partial RNG_{jnl}} = \frac{\beta_l \alpha_l \big( \ln RNG_{jnl} - \ln RNG_{jnl}^{ref} \big)^{\alpha_l - 1} + \sum_{r=1}^{R} \varphi_{rl}\, x_{nr}^*}{RNG_{jnl}} \tag{11}$$

## 5. Results and Discussion

We estimate three specifications of ICLV. In the first specification (Model 1), we consider a linear-in-utility specification and no interaction of alternative-specific attributes with latent variables (i.e., $\alpha_l = 1$ and $\varphi_{rl} = 0$ for all $r$ and $l$ in Equation 2). In the second specification (Model 2), we assume a prospect-theory-based non-linear specification, but do not consider interaction terms (i.e., $\varphi_{rl} = 0$ for all $r$ and $l$ in Equation 2). In Model 3, we estimate all model parameters.

We test different combinations of covariates to obtain the final specification, which can result in the highest goodness-of-fit and empirically as well as statistically significant parameter estimates. The specification consists of three latent variables – climate doubters, EV-tech believers, and early adopters. We name the latent variables based on the signs of their loadings ($\Pi$) on the observed ordinal indicators. We conduct the specification search for Model 3 and use the same covariates in Models 1 and 2.

We present parameter estimates of the DCM (i.e., binary Probit model) for all specifications in Table 5, but SEM parameter estimates are presented for only Model 3 in Tables 6 and 7 because the results of Models 1 and 2 provide similar insights. In terms of goodness-of-fit, as expected, Model 3 is better than Model 2, and both outperform Model 1 due to the inherent nesting structure (see Table 5). In this section, we discuss results of the binary Probit model, followed by the results of the SEM component of ICLV, WTP estimates, a brief comparison of WTP estimates with those reported by previous studies, and policy implications.

### 5.1 Discrete Choice Model (DCM)

Among all the product- and service-specific attributes presented in the choice experiment, we do not find a statistically significant effect of slow charging time, density of fast-charging stations, and privilege-driven policies on the preferences of Indian consumers for EVs (see Table 5). However, we have enough statistical evidence that purchase price, driving range, operating cost, and fast charging time determines the preferences of Indian car buyers for EVs. The parameter estimates of DCM in Table 5 indicate that the directions of the relationships between covariates and preferences are as expected. The results of Models 2 and 3 show that the curvature parameter $\alpha$ is less than 1 for all attributes, except for the fast charging time. This finding provides evidence of the existence of *reference dependence*, which have not been considered by previous studies on elicitation of consumer preferences for EVs. It is worth noting that intercepts of Model 1 are not directly comparable with those of Models 2 and 3 because of the non-linear specification, but WTP estimates are comparable. Therefore, we will detail the consequences of incorporating *reference dependence* while discussing the WTP estimates in Section 5.3.

The increase in the magnitude of indicator parameters $\vartheta$ with the increase in ordinal levels shows that higher knowledge of vehicle buyers about EVs and inclination of friends towards buying EVs positively affect their likelihood to purchase EVs. In all specifications, as expected, climate doubters and early adopters are less and more inclined to buy EVs, respectively, while keeping everything else constant. The results of Model 3 further exhibit that preference of climate doubters for EVs is negatively affected as the EV price deviates from the ICEV price. Interestingly, interactions between the latent variable for early adopters and other vehicle attributes indicate that early adopters care about operating cost and driving range more, but they are willing to buy an EV at slightly higher purchase price than ICEVs (see Model 3 in Table 5). Models 1 and 2 suggest that EV-tech believer are less likely to buy EVs, which at first appears counter-intuitive to some extent. However, after interacting the latent variable for EV-tech believers with the purchase price and the driving range in Model 3, we find positive parameter on the latent variable, negative parameter on its interaction with the purchase price, and positive parameter on its interaction with driving range. These relationships imply that if the price of EV is the same as ICEV, EV-tech believers would prefer to buy EVs over ICEVs. Their inclination to buy EVs would strengthen with the increase in the driving range of EVs and would weaken with the deviation of the EV price relative to the ICEV price.

**TABLE 5** Results of the discrete choice model (Equation 2, All models, N= 1021).

| Explanatory variables | Model 1 | Model 2 | Model 3 |
|---|---|---|---|
| **Electric vehicle (EV) utility** | | | |
| | *Alternative-specific attributes* {$\beta, \alpha$} | | |
| Constant | 0.17 | 1.85 | 1.88 |
| Fast charging time (in minutes) | -0.002 | -0.003 | -0.002 |
| EV price- ICEV price (in lacs) | | | |
| Intercept | -0.10 | -1.37 | -1.22 |
| Curvature | 1.00 | 0.24 | 0.31 |
| Log [EV range] - Log [ICEV range] (Range is in 100 Kilometres) | | | |
| Intercept | 0.60 | 0.74 | 0.94 |
| Curvature | 1.00 | 0.51 | 0.55 |
| ICEV weekly fuel cost - EV weekly fuel cost (fuel cost is in INR 100) | | | |
| Intercept | 0.009 | 0.020 | 0.024 |
| Curvature | 1.00 | 0.80 | 0.81 |
| | *Indicators* {$\vartheta$} | | |
| Ind10: Knowledge about EVs (Base: Have little knowledge) | | | |
| Never heard | -0.36 | -0.42 | -0.37 |
| Have heard, but no knowledge | -0.10 | -0.10 | -0.09 |
| Have a fair amount for knowledge | 0.22 | 0.22 | 0.24 |
| Know all about EVs | 0.49 | 0.49 | 0.51 |
| Ind11: If my friend buys EV, my chance of buying it will increase (Base: Neutral) | | | |
| Strongly disagree | -0.46 | -0.45 | -0.44 |
| Disagree | -0.32 | -0.32 | -0.31 |
| Agree | 0.69 | 0.70 | 0.73 |
| Strongly agree | 1.18 | 1.18 | 1.22 |
| | *Latent variables* {$\delta$} | | |
| Climate doubters | -0.61 | -0.61 | -0.55 |
| EV-tech believers | -0.22 | -0.21 | 0.10 |
| Early adopters | 0.94 | 0.93 | 0.55 |
| | *Interaction* {$\varphi$} | | |
| Climate doubters x [EV price- ICEV price] (price is in lacs) | | | -0.026 |
| EV-tech believers x [EV price- ICEV price] (price is in lacs) | | | -0.005* |
| Early adopters x [EV price- ICEV price] (price is in lacs) | | | 0.047 |
| EV-tech believers x Log [EV range / ICEV range] (range is in 100 kilometres) | | | 0.578 |
| Early adopters x Log [EV range] (range is in 100 kilometres) | | | 0.145 |
| Early adopters x EV weekly fuel cost (fuel cost is in INR 100) | | | -0.030 |
| **Internal combustion engine vehicle (ICEV) utility** | | | |
| | *Alternative-specific attributes* {$\beta$} | | |
| Fast charging time (in minutes) | -0.006 | -0.007 | -0.004 |
| | *Interaction* {$\varphi$} | | |
| Early adopters x ICEV price (in lacs) | | | -0.017 |
| Early adopters x Fast charging time (in minutes) | | | -0.010 |
| Loglikelihood | -3462.8 | -3457.2 | -3367.0 |

**Note:** Parameters with superscript * are statistically significant at 0.1 significance level, and all other parameters are statistically significant at 0.05 significance level.

## 5.2 Structural Equation Model (SEM)

Table 6 presents the parameter estimates of the structural part of the SEM (i.e., trivariate linear regression), where we also introduce location-specific fixed effects. When comparing with male vehicle buyers, females on an average have a stronger belief that EVs can help in the battle against climate change, and perhaps that is why they are more likely to be early adopters even after having relatively weaker trust in EV technology. Married families who have an annual household income below 20 lacs (~US$27000) are less firm regarding the contribution of EVs in addressing climate change issues as compared to their demographic counterparts. Moreover, despite having a lower trust in EV technology, they are more likely to be early adopters, perhaps they value the EV's future operating cost saving more than their counterparts (as also indicated by a statistically significant interaction of operating cost with the latent variable for early adopters in the binary Probit model).

**TABLE 6** Results of the structural equation of the structural equation model (Equation 4, Model 3, N= 1021).

| Explanatory variables | Climate doubters | EV-tech believers | Early adopters |
|---|---|---|---|
| | *Slope parameters {Π}* | | |
| *Respondent's location (Base: Delhi and others)* | | | |
| Mumbai | 0.027 | 0.013 | 0.080 |
| Bangalore | 0.259 | 0.168 | 0.094 |
| Chennai | 0.195 | -0.441 | -0.258 |
| Calcutta | -0.208 | 0.284 | 0.235 |
| *Gender (Base: Male)* | | | |
| Female | -0.050 | -0.230 | 0.188 |
| *Marital status (Base: Single and others)* | | | |
| Couple | 0.163 | -0.146 | 0.131 |
| Couple with kid | 0.113 | -0.254 | 0.115 |
| *Annual household income (Base: 20 lacs or more)* | | | |
| Less than 5 Lakh | 0.018 | -0.150 | 0.606 |
| 5-10 Lakh | 0.018 | -0.079 | 0.419 |
| 10-15 Lakh | 0.018 | -0.079 | 0.245 |
| 15-20 Lakh | 0.018 | -0.079 | |
| *Education level (Base: Master's degree or above)* | | | |
| Below Bachelor's degree | 0.025 | -0.297 | |
| Bachelor's degree | 0.025 | -0.151 | |
| *Employment type (Base: Private Sector)* | | | |
| Government employee | -0.192 | -0.572 | -0.529 |
| Self-employed | -0.242 | 0.168 | 0.281 |
| Unemployed | 0.248 | -0.066 | -0.228 |
| | *Error correlations {L$_\gamma$}* | | |
| Climate doubters vs. EV-tech believers | | -0.065 | |
| Early adopters vs. EV-tech believers | | 0.802 | |

**Note:** All parameter estimates are statistically significant at .01 significance level.

As expected, vehicle buyers with a bachelor's degree or below on an average have lesser trust in EV technology and abilities of EVs in addressing climate change issues as compared to those with higher education. We also observe diversity in preferences of vehicle buyers relative to employment status. For instance, unemployed respondents are not only less confident about EV technology and its impact on the environment, they are also less likely to adopt EVs early as compared to their counterparts who are working in the private sector. We also present estimates of the statistically significant elements of the error correlation matrix $L_\gamma$ in Table 6. A high error correlation of 0.8 between EV-tech believers and early adopters indicate that many unobserved factors in both latent variable equations are common.

Table 7 presents the parameters estimates of the measurement component of SEM, i.e. multivariate ordered Probit models. We observe that latent variables have statistically significant loading on corresponding ordinal indicators. All intercepts and thresholds (or cut-offs) are also statistically significant. Substantial differences between threshold estimates show that respondents could distinguish five ordinal levels for all ordinal indicators. All the latent variables have statistically significant loading on knowledge (Ind10) and social network effect (Ind11) indicators. The results indicate that those who believe more in EV technology and the significance of EVs in battling against climate change but do not want to adopt EVs early, are likely to have a higher awareness about EVs. Moreover, those who have a better impression about abilities of EVs to mitigate climate change and are likely to adopt EVs at low market penetration, even after having lower trust in EV technology, are prone to experience higher social network effects in EV purchase decisions.

**TABLE 7** Results of measurement equation of the structural equation model (Equations 5 to 8, Model 3, N= 1021).

| Indicator statements on Likert scale (strongly disagree = 1, strongly agree = 5, if not specified) | Climate doubters | EV-tech believers | Early adopters | Intercept | Threshold 2 | Threshold 3 | Threshold 4 |
|---|---|---|---|---|---|---|---|
| Ind01: We can reduce climate hazard by changing our behaviour. | -0.94 | | | 2.64 | 0.33 | 1.13 | 2.79 |
| Ind02: I am concerned about the influence of human behaviour on climate change. | -0.74 | | | 2.73 | 0.38 | 1.30 | 3.20 |
| Ind03: EVs can help in battling climate change. | -0.96 | | | 2.97 | 0.63 | 1.61 | 3.31 |
| Ind04: I do not want to take a risk by purchasing an EV since I know little about it. | | -1.05 | | 1.55 | 1.73 | 2.69 | 4.28 |
| Ind05: I do not trust EV technology. | | -0.87 | | 1.80 | 1.42 | 2.32 | 3.90 |
| Ind06: EV is not a proven technology, which discourages me from purchasing. | | -0.86 | | 1.78 | 1.42 | 2.26 | 3.77 |
| Ind07: Limited service support for EVs in India discourages me from purchasing. | | | -0.69 | 2.74 | 0.84 | 1.67 | 3.23 |
| Ind08: Not a good idea to purchase an EV due to the limited charging infrastructure. | | | -0.75 | 3.00 | 1.28 | 2.02 | 3.67 |
| Ind09: It is not a good idea to purchase an EV due to low resale value. | | | -0.44 | 1.99 | 1.19 | 2.12 | 3.31 |
| Ind10: Knowledge about EVs (1= never heard to 5 = know all). | -0.08 | 1.33 | -1.32 | 3.86 | 1.79 | 3.33 | 5.29 |
| Ind11: If my friend buys EV, my chance of buying it will increase. | -0.38 | -0.22 | 0.03 | 1.55 | 0.83 | 1.78 | 3.19 |

**Note:** All parameter estimates are statistically significant at .01 significance level.

## 5.3 Willingness to Pay (WTP)

We input parameter estimates of ICLV in Equations 9 to 11 to estimate WTP of Indian consumers to reduce fast charging time by 10 minutes, improve driving range by 100 kilometres, and reduce the weekly operating cost of EVs by INR100 (~US$1.4). It is worth noting that WTP estimates in prospect-theory-based models (i.e., Model 2 and Model 3) depend on both EV and ICEV (reference alternative) attribute values on which marginal changes in attributes occur. Such dependence brings more realism in WTP estimates as compared to those obtained using traditional linear-in-parameter utility specification (Mabit et al., 2015).

It is infeasible to present WTP results for all possible attribute values of the reference alternative (i.e., ICEV). Therefore, we choose charging time, driving range, weekly operating cost, and purchase price of ICEV to be 5 minutes, 800 kilometres, INR 500 (~US$6.7), and INR 10 lacs (~US$13500), respectively, in all computations. Moreover, it is important to realise that WTP estimates from Model 3 depend on the values of latent variables due to interactions effects, which in turn are a function of an individual's demographic characteristics. Thus, Model 3 accounts for heterogeneity in preferences of various demographic groups. Considering practical aspects, we present WTP estimates from Model 3 for only two very different demographic groups and EV price of INR 13 lacs (~US$17,350). First, "demographics 1" indicates a male vehicle buyer who is married and unemployed, has an annual household income below INR 5 lacs (~US$6750), has education below bachelor's degree, and lives in Bangalore. Second, "demographics 2" represents a self-employed single female whose annual household income is INR 20 lacs (~US$27000) or above, has a master's degree or above, and lives in Calcutta. We plot WTP estimates in Figure 4 for all models and discuss them below in detail. These plots are illustrative; other plots can be generated for different attributes of the reference alternative and other demographic groups using estimates of Model 3.

Plot (a) in Figure 4 shows estimates of WTP to reduce the EV fast charging time by 10 minutes. Model 1 estimates that Indian consumers are willing to pay additional INR 20,900 (~US$279) to reduce EV fast charging time by 10 minutes. Since we do not observe *diminishing sensitivity* of marginal utility for fast charging time, WTP estimates do not vary at different values of fast charging time in Models 2 and 3. However, they do vary with changes in EV price because *diminishing sensitivity* exists for the purchase price. For instance, WTP estimates from Model 2 to

reduce fast charging time by 10 minutes are INR 7.5, 17.2, and 25.3 thousand (~US$100, US$230, and US$337) for EV price of INR 11 lacs, 13 lacs, and 15 lacs (~US$14,680, US$17,350, US$20,000), respectively. Instead of providing a single estimate of WTP using traditional linear-in-parameter utility specification, the prospect-theory-based specification could offer more insights about the variation in WTP at different attribute levels. For EV price of 13 lacs, Model 3 estimates that WTP to reduce fast charging time by 10 minutes for "*demographics 1*" and "*demographics 2*" are INR 14.4 and 17.9 thousand (~US$192 and US$239), respectively.

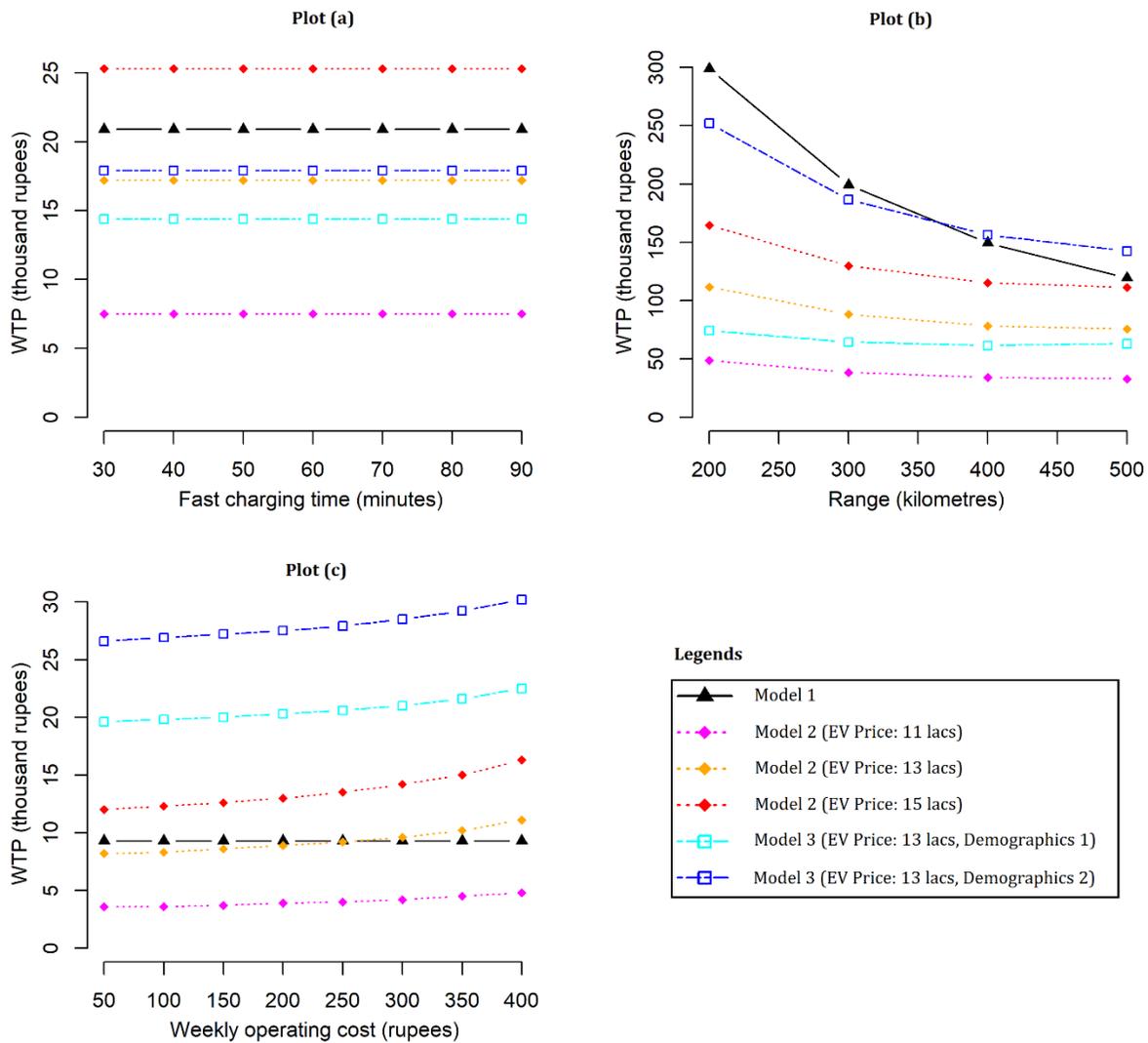

**FIGURE 4** Marginal willingness to Pay (WTP) of Indians for (a) reduction in fast charging time of EV by 10 minutes (ICEV refueling time: 5 minutes); (b) increasing driving range of EV by 100 kilometres (ICEV driving range: 800 kilometres); (c) reducing weekly operating cost of EV by INR 100, i.e. ~US$1.4 (ICEV weekly operating cost: INR 500, i.e. ~US$6.7). In all plots, ICEV price is INR 10 lacs (~US$13500). Note that INR 1000 are equivalent to ~US$13.3.

Plot (b) in Figure 4 shows WTP estimates to improve the EV driving range by 100 kilometres. As a consequence of the logarithmic transformation of driving range in all specifications, WTP to increase EV driving range decreases with the EV driving range. This trend is aligned with the intuition because the utility gain due to marginal increase in driving range at a lower driving range value is higher. For instance, Model 1 shows that Indian consumers are willing to pay an additional INR 298 thousand (~US$3,977) to increase the EV driving range from 200 to 300 kilometres, but the WTP reduces to INR 119 thousand (~US$1,588) to improve the range from 500 to 600 kilometres. These estimates from Model 2 are INR 111 thousand and 75.6 thousand (~US$1480 and US$1010) when EV price is INR 13 lacs (~US$17,350), which change to INR 74.3 thousand and 63.1 thousand (~US$991 and US$842) for "*demographics 1*", and INR 252 and 142 thousand (~US$3363 and US$1895) for "*demographics 2*" in Model 3. Similar to the WTP estimates to reduce fast charging time, estimates of WTP to improve driving range from Model 2 lie between those obtained from Model 3 for "*demographic 1*" and "*demographic 2*". The results indicate that imposing logarithmic transformation in linear-in-utility specification leads to sharp (and structural) decline in the WTP estimates with the increase in the driving range, but the estimable curvature in the prospsect-theory-based utility specification better moderates this decrease trend in a data-driven manner.

Plot (c) in Figure 4 shows WTP estimates to reduce weekly operating cost by INR 100 (~US$1.4). According to Model 1, Indian consumers are willing to pay only additional INR 9.3 thousand (~US$125) in the purchase price to reduce the weekly operating cost of an EV by INR 100 (~US$1.4). When EV price is INR 13 lacs (~US$17,350), the WTP estimate for Model 2 increases from INR 8.2 thousand to 11.1 thousand (~US$110 to US$148) as EV operating cost increases from INR 50 (~US$0.7) to INR 400 (~US$5.7) per week. These numbers are slightly higher for Model 3 – INR 19.6 thousand and 22.5 thousand (~US$261 and US$300) for "*demographics 1*", and INR 26.6 thousand and 30.2 thousand (~US$355 and US$403) for "*demographics 2*". This trend is a consequence of the higher magnitude of the parameter of the interaction between the weekly fuel cost and the *early adopter* latent variable in Model 3.

To put these WTP estimates in perspective, we compute the annual discount rate that consumers consider to compute the present value of the future fuel cost savings before compensating these savings with the purchase price. To this end, we use the present value of annuity formula

$A \times \left[\frac{1-(1+i)^{-n}}{i}\right] = P$, where $A$ is the weekly fuel cost savings (INR 100, i.e. ~US$1.4 in our case), $i$ is the weekly discount rate, $n$ is the number of weeks in the life span of 15 years ($n = 15\ years \times 52\ weeks = 780$), and $P$ is the WTP to reduce the weekly fuel cost by INR 100 (~US$1.4) (Sullivan et al., 2003). We then convert the weekly discount rate into the annual discount rate $r$ using $r = (1+i)^{52} - 1$. Corresponding to WTP estimates of INR 9.3 thousand (~US$125) from Model 1, the annual discount rate is 74.3%. However, the annual discount rates corresponding to INR 26.6 thousand and 30.2 thousand (~US$355 and US$403) for "*demographics 2*" in Model 3 are 20.0% and 16.7%, respectively. Given that higher discount rate implies lower valuation of future fuel cost savings, these estimates indicate that linear-in-utility parameter specification overestimates the consumers' undervaluation of the fuel economy. Given that the market interest rate is around 4% (Economics, 2020), even discount rate estimates of Model 3 indicate that Indian car buyers are somewhat myopic and associate lower value to the future fuel cost savings. Previous studies have also got similar results regarding car buyers' fuel economy valuation (Turrentine and Kurani, 2007).

## 5.4 Comparison of WTP estimates

The comparison of WTP estimates across studies is inherently challenging because most studies estimate WTP for different "amount of change" in attributes (e.g., some present WTP for the increase in driving range by 1 km and others for 100 km increment). These differences become a concern in comparative analysis because the WTP cannot be linearly extrapolated or interpolated for many attributes due to non-linear utility specification (e.g., driving range always enters as logarithm because marginal WTP should decrease with the increase in driving range). The inclusion of taste heterogeneity makes this comparison even more cumbersome because a range of WTP values (instead of the point estimate) is generally provided. Moreover, differences in the current level of EV adoption also create heterogeneity in WTP estimates across different geographies. Nonetheless, we observe that our WTP estimates are broadly in line with those reported by previous studies.

WTP of Indian car buyers is US$7-40 to add one kilometre in driving range of EVs at the driving range of 200 kilometres, which is similar to or slightly lower than the range of WTP estimates reported for other countries. For example, Hoen and Koetse (2014), Parsons et al. (2014),

Hackbarth and Madlener (2013), Jensen et al. (2013), and Hess et al. (2012) find that consumers are willing to pay an additional upfront price of US$63 (Netherlands), US$33–71 (USA), US$10–40 (Germany), and US$20–235 (Denmark), and US$25–92 (California, USA), respectively, to add one-kilometre in driving range of EVs. Among more recent studies, Hackbarth and Madlener (2016) estimate that German's marginal WTP for additional kilometre of EV driving range is US$22-422 with the mean value of US$116 at the driving range of 100 kilometres and US$5-84 with the mean value of US$23 at the driving range of 500 kilometres. In another study, Huang and Qian (2018) find that the marginal WTP of Chinese consumers is around US$75 for an additional kilometre of driving range.

Whereas Indian consumers are willing to pay US$10-34 to reduce the fast charging time by a minute, German consumers are willing to pay US$9–76 with the mean value of US$25 when the EV charging time is one hour (Hackbarth and Madlener, 2016). Similarly, by applying LC-MNL on the stated preference data of over 17000 Canadian vehicle buyers, Ferguson et al. (2018) find that Canadians are willing to pay around US$33 more in the purchase price to save a minute of public charging time.

Using the average weekly kilometres travelled (230 kilometres) of the sample, we convert WTP estimates to reduce operating cost in a comparable unit. We find that Indian consumers are willing to pay an additional US$104-692 in the purchase price to save US$1 per 100 kilometres. Similarly, Jensen et al. (2013) find that Danish consumers are willing to pay US$79–200 to save US$1 per 100 kilometres in operating cost. According to the findings of Huang and Qian (2018), Chinese consumers are willing to pay approximately US$3900 more as the upfront cost to save US$1550 annually in operational cost. If we convert our WTP estimates on the same scale, Indian consumers are willing to pay around US$1340-8926 to save US$1550 annually in operational cost.

**5.5 Policy Implications**

By quantifying the effect of the product, service, policy, and attitudinal factors on preferences of Indian consumers for EVs, this study offers several insights – a) key barriers to the adoption of EVs, b) devising informed marketing strategies, c) efficient allocation of investment in research and development, and d) targeted EV design and optimal prices. We succinctly discuss how all these insights would help policymakers and automakers in creating an eco-system to accelerate the adoption of EVs in India.

The sample summary statistics show that a large proportion of Indians are concerned about the environment, but very few of them know about EVs. Moreover, the results of the SEM component of ICLV identify the demographic groups, who have doubts about the ability of EVs in battling climate change, have low trust in EV technology and are early adopters. Whereas summary statistics highlight the need for an extensive EV awareness campaign in India, the SEM results could assist policymakers in the effective design of such a campaign. For instance, one can identify regions with a higher proportion of early adopters and can customise the advertising strategy accordingly.

These findings are also relevant from an automaker's perspective. The strong presence of social norm/network effect among Indian car buyers, also observed by Schuitema et al. (2013) and White and Sintov (2017), implies that targeting early adopters can be an efficient method for EV diffusion in new markets. This strategy would be viable in the long run because word-of-mouth can further drive the mass adoption of EVs, as also anticipated by Mercedes-Benz US chief (Stankiewicz, K., 2019). Our SEM results can help automakers in targeting early adopters. For instance, based on our SEM finding that females are likely to adopt EVs earlier than males, also corroborated by Hidrue et al., (2011) and Ghasri et al. (2019), automakers might want to customise some EV models with lucrative features for females. Some of the female-friendly features in the Indian context are automatic transmission, rear parking cameras, assisted navigation, and electronically foldable mirrors (Mukherjee et al., 2020). Such inclusive designs would not require a paradigm shift in automaker's strategy as the anecdotal evidence suggest that automakers have already been targeting females to increase the sales – the share of female car buyers has increased from 7% to 12% in last five years, and it is expected to increase further (Mukherjee et al., 2020). Moreover, while designing an EV variant of an ICEV, demographic-specific WTP estimates can assist automakers in selecting the EV-specific features to ensure that the existing ICEV adopters do not find the EV variant too expensive.

Among product and service attributes, driving range and fast charging time are the key determinants of EV adoption by Indian consumers. In this regard, the Indian government has already taken multiple initiatives to minimise the range anxiety of drivers by improving the charging infrastructure. For instance, FAME-II scheme allocates INR 10 billion (~US$130 million) for development of charging stations with incentives from 50-100% on the cost of charger based on the access and locations (Government of India, 2019b). In addition, Indian consumer's undervaluation of future operating cost savings implies that policies like fuel tax might not be effective in expediting the EV adoption in India.

In contrast to the findings of the previous studies, our results suggest that privilege-driven policies such as specialised lanes and reserved parking for EVs might not be effective in accelerating the early adoption of EVs in India. Perhaps, Indian consumers are more concerned about product and service attributes of EVs in the first place. Such privilege-driven policies might be successful at slightly later stages – for example, at 5% market penetration of EVs. Our results also flag a warning for planned privilege-driven policies. For instance, Delhi state government should be cautious before reserving 20% of all residential and official parking spaces for EVs (Singh, 2019). Other government agencies should also understand the perception of locals before implementing such policies.

## 6. Conclusions

Using data on stated preferences of over 1000 Indian consumers, we present the first estimates of WTP of India consumers for EV attributes (e.g., driving range and fast charging time). To simultaneously understand the effect of latent attitudes (i.e., environment-friendliness and social norms) on the likelihood of Indian car buyers to adopt EVs, our analysis relies on a hybrid choice model. In this framework, we also account for the *reference dependence*. The results indicate that instead of providing just a single WTP estimate, *reference dependence* provides more realistic WTP estimate by allowing them to vary based on attributes of the reference alternative (ICEV, in our case). By interacting latent variables with vehicle attributes, we also capture observed preference heterogeneity in WTP estimates. Our results show that Indian consumers are willing to pay an additional US$10-34 in the purchase price to reduce the EV fast-charging time by one minute, US$7-40 to add a kilometre to EV driving range at 200 kilometres, and US$104-692 to save US$1 per 100 kilometres in future operating cost. These estimates are aligned with those reported in previous studies.

Since this is the first such study in the Indian context, several avenues for future research emerge. First, the results of this study cannot be generalised to the entire nation because the representativeness of the collected sample cannot be tested in the absence of demographic characteristics of the target population (car owners/buyers). The entire Indian census data cannot be used to compute demographic characteristics of the target population because car buyer and non-buyers differ substantially in terms of income and education levels. Future effort can be focused on acquiring disaggregate Indian census data to create the demographic distribution of car buyers in India. Subsequently, future studies can either collect data such that the joint demographic distribution of the sample matches with that of the population or can use sampling weights to make the sample representative of the population. Second, considering the diversity of India, validating the estimates of this study on a larger sample is an essential next step. To avoid the excessive cognitive burden on respondents, our choice experiment design does not account for the vehicle body type, emissions, and performance attributes (e.g., acceleration and top speed). The subsequent survey can potentially include these attributes. Third, the current study only focuses on consumer demand for EVs. Accounting for interaction between demand and supply sides would offer more insights about the market-level substitution effects (Berry et al., 1995). These equilibrium models could also assist policymakers in designing policies to increase the EV adoption by simulating the market-level impact of policies. Such market-level analysis is even more relevant in the Indian context as the Indian Government is envisioning feebate policies as a policy lever to accelerate the EV adoption (NITI Aayog and Rocky Mountain Institute, 2017b).